\documentclass[preprint,12pt, nofootinbib]{revtex4}
\usepackage[colorlinks=true, pdfstartview=FitV, linkcolor=blue, citecolor=red, urlcolor=magenta]{hyperref}
\usepackage[brazil, english]{babel}
\usepackage[utf8]{inputenc}
\usepackage{graphicx}
\usepackage{latexsym}
\usepackage{amsmath}
\usepackage{amsfonts}
\usepackage{amssymb}
\usepackage{verbatim}

\usepackage{tikz,tikz-3dplot}
\tdplotsetmaincoords{80}{45}

\tikzset{surface/.style={draw=black, fill=white, fill opacity=.6}}

\usepackage{tikz}
\usetikzlibrary{decorations.pathmorphing,patterns}

\newcommand{\be}{\begin{equation}}
\newcommand{\ee}{\end{equation}}
\newcommand{\bea}{\begin{eqnarray}}
\newcommand{\eea}{\end{eqnarray}}

\newcommand{\ben}{\begin{eqnarray}}
\newcommand{\een}{\end{eqnarray}}

\graphicspath{%
    {converted_graphics/}
    {/}
}
\begin{document}

\title{AdS/BCFT correspondence and BTZ black hole within electric field}

\author{Fabiano F. Santos$^{a}$}
\email{fabiano.ffs23-at-gmail.com}\affiliation{$^{a}$Instituto de Física,Universidade Federal do Rio de Janeiro, Caixa Postal 68528, Rio de Janeiro-RJ, 21941-972 -- Brazil}

\begin{abstract}
This paper, presents the gravity duals of Conformal Field Theories with boundaries. This theory is known as AdS/BCFT correspondence. In this duality, our system has 3D gravity coupling with the Maxwell field dual to 2D BCFT. On the gravity side, we consider a 3D BTZ black hole. We analyze the effects of the chemical potential on the profile of the extra boundary for the black hole. Performing a holographic renormalization, we calculate the free energy and obtain the total entropy and corresponding area with chemical potential, and the boundary entropy for the black hole. These theories are specified by stress-energy tensors that reside on the extensions of the boundary to the bulk. In this way, the AdS/BCFT appears analogous to the fluid/gravity correspondence with the chemical potential. We discuss the solutions as well as their thermodynamic and fluid properties.
\end{abstract}
\maketitle
\section{Introduction}

\begin{figure}[!htbp]
    \centering
    \scalebox{1.4}{
    \pgfkeys{/tikz/.cd, view angle/.initial=0, view angle/.store in=\picangle}
\tikzset{
  horizontal/.style={y={(0,sin(\picangle))}},
  vertical at/.style={x={([horizontal] #1:1)}, y={(0,cos(\picangle)cm)}},
  every label/.style={font=\tiny, inner sep=1pt},
  shorten/.style={shorten <=#1, shorten >=#1},
  shorten/.default=3pt,
  ->-/.style={decoration={markings, mark=at position #1 with {\arrow{>}}}, postaction={decorate}},
  ->-/.default=0.5
}
\begin{tikzpicture}[scale=1.34, view angle=15, >=stealth]
  \draw[fill=green!50, fill opacity=.2, horizontal, thick] (0,0) circle (1);
 (0:-105:1);
  \draw[fill=red, fill opacity=.3, color=red, thick] (0:1) arc (0:-180:1) [horizontal] arc (-180:0:1);
  \draw[fill=gray!50, fill opacity=.0, horizontal, color = black, thick] (0,0) circle (1);
    \draw[fill=red, fill opacity=.0, color=black, thick] (0:1) arc (0:-180:1) [horizontal] arc (-180:0:1);
 (0:-105:1);

    \node[] at (0,0) 
    { $\mathcal{M}_{\mathrm{AdS}} $};

    \node[] at (0,-0.7) 
    {$\mathrm{CFT}$};
    \node[] at (1.1,-0.55) 
    {$Q$};
\end{tikzpicture}
}
    \caption{AdS/CFT correspondence in the presence of boundary hypersurface $Q$}
    \label{p0}
\end{figure}
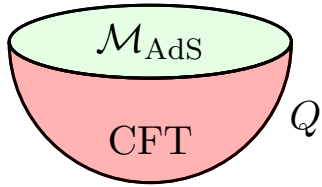

Several considerations around the AdS/CFT correspondence have received increasing attention recently  \cite{Jiang:2017imk,Santos:2022oyo,Baggioli:2017ojd,Li:2018kqp,Li:2018rgn,Hartnoll:2009sz,Hartnoll:2008hs,Lucas:2015vna,Sachdev:2011wg,Kovtun:2003wp,Kovtun:2004de,Feng:2015oea,Sadeghi:2018vrf,	Brito:2019ose,Baggioli:2016rdj,Liu:2018hzo}, this theory relates to a Yang-Mills theory $SU(N_{c})$ in $\Re^{1,3}$ with large $N_{c}$ and supersymmetry $\aleph=4$ for a Superstring Theory or Supergravity in AdS$_{5}$xS$^{5}$ where AdS is the Anti-de-Sitter spacetime while CFT means Conformal Field Theory \cite{Maldacena:1997re,Gubser:1998bc,Witten:1998qj,Aharony:1999ti}. The AdS/CFT correspondence \cite{Hartnoll:2009sz,Hartnoll:2008hs,Lucas:2015vna,Sachdev:2011wg} provides universal bounds of the transport coefficients, including the well-known bound ratio of the shear viscosity \cite{Kovtun:2003wp,Kovtun:2004de}. All these bounds  are conjectured based on the holographic ``bottom-up'' models, which are violated through ways put forward in \cite{Feng:2015oea,Sadeghi:2018vrf,Brito:2019ose,Baggioli:2016rdj,Liu:2018hzo}. However, just as the AdS/CFT correspondence is used to study the transport coefficients, a boundary theory known as AdS/BCFT correspondence has recently emerged \cite{Takayanagi:2011zk,Fujita:2011fp}, which is an extension of the AdS/CFT correspondence \cite{Fujita:2012fp,Melnikov:2012tb,Cavalcanti:2018pta}. The idea behind the AdS/BCFT correspondence proposed by Takayanagi \cite{Takayanagi:2011zk} consists of an extension of the AdS/CFT correspondence \cite{Maldacena:1997re}, defining in the CFT a boundary in $d$-dimensional manifold $\mathcal{M}$ for a $d+1$-dimensional asymptotically AdS space so that $\partial N=\mathcal{M}\cup Q$, where $Q$ is a $d$-dimensional manifold that satisfies $\partial Q=\partial \mathcal{M}=P$ figure (\ref{p0}). This new extension has brought attention to investigating the transport coefficients like the Hall effect, fluid/gravity correspondence and the Hawking-Page phase transition \cite{Takayanagi:2011zk,Fujita:2011fp,Melnikov:2012tb,Cavalcanti:2018pta,Santos:2021orr,Magan:2014dwa}. 


In our prescription, our gravitational system with a BTZ black hole coupled to the Maxwell field, we impose Neumann boundary conditions. For this system, we construct the AdS/BCFT, where we find the location of $Q$ in the bulk with an electric field. In $(2+1)$-dimensional, gravity is considered a toy model \cite{Carlip:1995qv}. It has neither a Newtonian limit nor any  propagating degrees of freedom. On the other hand, the work of Bañados, Teitelboim, and Zanelli (BTZ) \cite{Banados:1992wn,Banados:1992gq} has a solution in (2+1) known as BTZ black hole, which has some interesting features, such as: an event horizon (and in some cases, an additional inner horizon, if one includes rotations)  presenting thermodynamic properties somehow similar to the black holes in (3+1) dimensions and being asymptotically anti-de Sitter. Usually, these black holes have asymptotically flat behavior. In our prescription, we choose to work with a planar BTZ black hole with a nontrivial axis profile. However, the Maxwell fields are given in terms of the 'electromagnetic duality' where $\mathcal{F}_{\mu\nu}$ is the Hodge dual of the field strength $^{*}\mathcal{F}_{\mu\nu}$ \cite{Miao:2018qkc,Miao:2018dvm,Miao:2017gyt,Miao:2017aba,Chu:2018ntx}. 

In our system, we compute the thermodynamics through the holographic renormalization scheme and analyze the thermodynamics of the BCFT$_{2}$. The thermodynamics of the BTZ black hole consists of volume and boundary contributions with electric field contributions. In this sense, the boundary hypersurface tension defines a characteristic scale for the corresponding thermodynamic quantities. In calculating the contribution of the boundary to the total action, we find that its associated entropy is not equal to the contribution from the Bekenstein-Hawking formula alone due to the presence of the electric field.

Within this fluid/gravity correspondence framework, \cite{Melnikov:2012tb,Cavalcanti:2018pta,Santos:2022oyo,Magan:2014dwa} we describe a family of boundary stress-energy tensors T$_{ab}$ that reside in $Q$. In our case, each T$_{ab}$ corresponds to a hypersurface in the volume that bounds a subspace of the BTZ black hole solution with an electric field. These hypersurfaces preserve all but one of the spatial translation symmetries. Therefore, from the point of view of the AdS/BCFT correspondence this is an AdS/CFT problem in half- space or an infinite range. In the fluid/gravity correspondence, the slices of constant radius of the AdS space (black hole) in the usual cases without additional fields lead to the stress-energy tensor of a conformal fluid, which only occurs in the low  temperature regime due to the presence of the electric field. In this work, we are interested in hypersurfaces that extend along the radial direction, i.e., from the boundary to the bulk. This external electric field causes the corresponding Tab not to assume the form of a simple fluid. In our results, we show that even with a unique geometry of the hypersurface that produces a fluid-like T$_{ab}$, we observe that both the energy density behavior decreases while the pressure behavior increases and, both behaviors are controlled by the electric field. Thus, in addition to the fluid in the boundary extension being subject to a curved metric, or equivalently to an external field, it will also be subject to an electric field.

In this work, motivated by the application of AdS/CFT duality together with the emergence of AdS/BCFT \cite{Takayanagi:2011zk,Fujita:2011fp,Santos:2021orr,Santos:2020xox} and taking into account the importance of ($2+1$)-dimensional black holes, we propose a description of the BTZ black hole in AdS$_{3}$/BCFT$_{2}$ correspondence with electric field \cite{Miao:2018qkc,Miao:2018dvm,Miao:2017gyt,Miao:2017aba,Sudhaker,Ren:2010ha}. The summary of the main results achieved are organized as:

\begin{itemize}
  
    \item First, we study the influence of the electric field on the BCFT theory. Apart from a complete numerical solution, we show a solution that is useful for determining the role of $Q$ profile and perform an analysis of all quantities in this work; 
    
    \item We construct a holographic renormalization for this setup and compute the free energy for the AdS-BTZ black hole;
    
    \item From the free energy, we compute the total and boundary entropy. In the case of the boundary entropy within an electric field, one can see it as an extension of the results found in \cite{Takayanagi:2011zk, Fujita:2011fp}; 
   
   \item In the fluid/gravity correspondence, the boundary extension is subject to a curved metric, or equivalently to an external field and electric field. As a result, the thermodynamic quantities, except entropy density, are coordinate dependent. Nevertheless, at every spatial position, the fluid has the same equation of state as the well-known conformal fluid defined at the corresponding energy scale in high temperatures. \cite{Takayanagi:2011zk, Fujita:2011fp,Magan:2014dwa}.  
\end{itemize}

This work is summarized as: In Sec.$\sim$\ref{v0} we present the setup AdS$_{3}$/BCFT$_{2}$. In Sec.$\sim$\ref{v1} we discuss the profile boundary conditions of the boundary $Q$ in terms of the gauge field.  In Sec.$\sim$\ref{v2} we present the entropy of the BTZ black hole within the electric field. In Sec.$\sim$\ref{v3} at the boundary fluid from AdS$_{3}$/BCFT$_{2}$ appears the fluid/gravity correspondence in terms of the electric field. Finally, in Sec.$\sim$\ref{v4}, we present our conclusions.

\section{The scenario AdS$_{3}$/BCFT$_{2}$}\label{v0}

The holographic dual from AdS$_3$/BCFT$_{2}$ is a simple example with multiple boundaries \cite{Takayanagi:2011zk,Fujita:2011fp,Fujita:2012fp,Melnikov:2012tb,Magan:2014dwa,Miao:2018qkc,Miao:2018dvm,Miao:2017gyt,Miao:2017aba,Chu:2018ntx,Ren:2010ha}. For that reason, we will use a holographic scenario considering in the AdS$_{3}$/BCFT$_{2}$ an electric field where the idea is to explore the boundary profile within the electric field. Thus, the action for this scenario is given by
\begin{eqnarray}
S=S_{N}+S_{Q}=\int_{N}{d^{3}x\sqrt{-g}\left((R-2\Lambda)-\frac{1}{4}\mathcal{F}_{\mu\nu}\mathcal{F}^{\mu\nu}\right)}+2\int_{Q}{d^{2}x\sqrt{-h}(K-\Sigma)},\label{1}
\end{eqnarray}
Where $K_{ab}h^{c}_{a}\nabla_{c}n_{b}$ is the extrinsic curvature, $h_{ab}$ is the induced metric, $K=h^{ab}K_{ab}$ is the traceless contraction, $n^{a}$ is the normal vector of the hypersurface $Q$ and $\Sigma$ are the tension of the boundary. Imposing a Neumann boundary condition in the action (\ref{1}), we have:

\begin{eqnarray}
K_{ab}=(K-\Sigma)h_{ab}.\label{2}
\end{eqnarray}
For more discussion about this procedure, see \cite{Takayanagi:2011zk,Fujita:2011fp,Fujita:2012fp,Melnikov:2012tb,Magan:2014dwa}. The Einstein-Maxwell equations can be formally written as: 
\begin{eqnarray}
&&G_{\mu\nu}+\Lambda g_{\mu\nu}=\left(\mathcal{F}_{\mu\alpha}\mathcal{F}^{\alpha}_{\nu}-\frac{g_{\mu\nu}}{4}\mathcal{F}_{\alpha\beta}\mathcal{F}^{\alpha\beta}\right)\label{4},\\
&&\nabla^{\mu}\mathcal{F}_{\mu\nu}=0.\label{4.1}
\end{eqnarray}
Now, to find the electric field black hole in three-dimension, we consider the following metric and gauge field, respectively, as:
\begin{eqnarray}
ds^{2}&=&\frac{L^{2}}{z^{2}}\left(-f(z)dt^{2}+dy^{2}+\frac{dz^{2}}{f(z)}\right),\label{3}\\
A&=&\phi(z)dt.\label{3.1}
\end{eqnarray}
Solving the Einstein-Maxwell equations for the metric (\ref{3}), the horizon function and the gauge field are given by:
\begin{eqnarray}
f(z)&=&1-\frac{z^{2}}{z^{2}_{h}}+\frac{\mu^{2}z^{2}}{L^{2}}\ln\left(\frac{z}{z_{h}}\right),\label{5}\\
A&=&\mu\ln\left(\frac{z}{z_{h}}\right)dt,\label{6}
\end{eqnarray}
These functions are solutions to the Einstein-Maxwell equations. Besides, the largest of the positive roots $z_{h}$ of the function $f(z)$–the so-called blackening factor–determines the horizon of the black hole. The black hole temperature is given by
\begin{eqnarray}
T=\frac{|f^{'}(z_{h})|}{4\pi}=\frac{4-\mu^{2}z^{2}_{h}}{8\pi z_{h}}.\label{TH}
\end{eqnarray}
Observing the equation (\ref{6}), we can extract the 'electromagnetic duality', which is written as
\begin{eqnarray}
\mathcal{F}_{\mu\nu}\to ^{*}\mathcal{F}_{\mu\nu}=\frac{1}{2}\epsilon_{\mu\nu\alpha\beta}\mathcal{F}^{\alpha\beta}\label{S3}.
\end{eqnarray}
Such duality, gives $\mathcal{F}_{\mu\nu}\mathcal{F}^{\mu\nu}=^{*}\mathcal{F}_{\mu\nu}^{*}\mathcal{F}^{\mu\nu}$ that says action (\ref{1}) is invariant under this duality (\ref{S3}) \cite{Miao:2018qkc}. Besides, the transformation (\ref{S3}) shows that $\mathcal{F}_{zt}\to ^{*}\mathcal{F}_{zt}=\mu/z$ where $\mu$ is the chemical potential and $\mathcal{F}_{zt}$ is the electric field, normal to the plane $\mathcal{M}$. We expect that for non-trivial values of the chemical potential, i.e., $\mu\neq 0$ the conformal symmetry must be broken.

\section{Q-profile within charged BTZ black hole}\label{v1}
\begin{figure}[!ht]
\begin{center}
\includegraphics[scale=0.5]{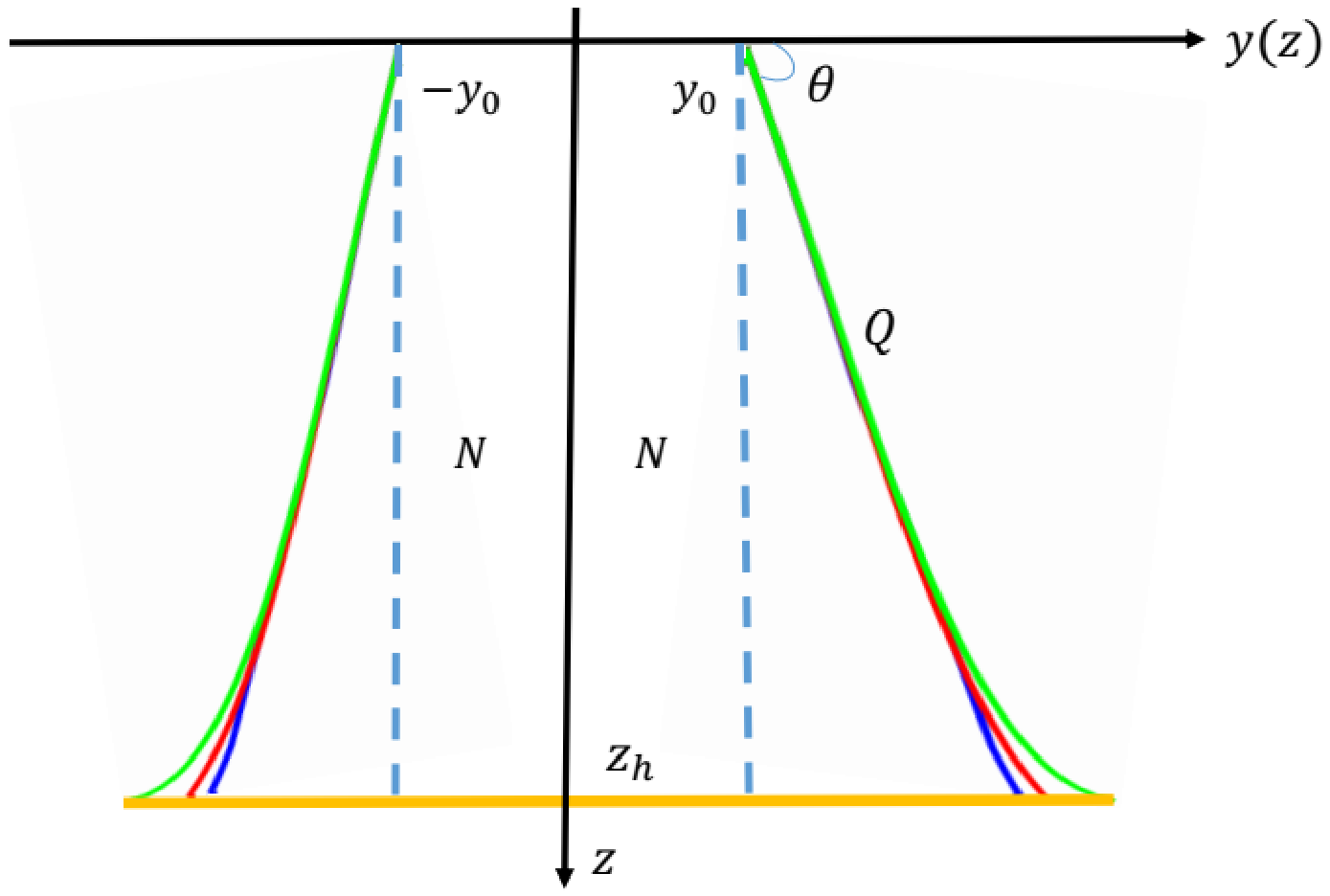}
\caption{$Q$ boundary profile for the BTZ black hole within the electric field for the values $\mu=1.12$ (blue curve), $\mu=1.14$ (red curve) and $\mu=1.16$ (green curve). Orange regions show the part of the horizon (shadows of $Q$), which contributes to the boundary entropy.}\label{p1}
\label{planohwkhz}
\end{center}
\end{figure}

Now, to construct the Q boundary profile, one has that the induced metric on this surface given by 
\begin{eqnarray}
ds^{2}_{\rm ind}=\frac{L^{2}}{z^{2}}\left(-f(z)dt^{2}+\frac{g^{2}(z)dz^{2}}{f(z)}\right)\,\label{metricindu}, 
\end{eqnarray}
where $g^{2}(z)=1+{y'}^{2}(z)f(z)$ with $y{'}(z)=dy/dz$. The normal vectors on Q are: 
\begin{eqnarray}
n^{a}=\frac{z}{Lg(z)}\, \left(0,\, 1, \, -{f(z)y{'}(z)}\right)\,.
\end{eqnarray}
The surface $Q$ is described by the curve $y=y(z)$ \cite{Takayanagi:2011zk,Fujita:2011fp,Fujita:2012fp,Melnikov:2012tb,Magan:2014dwa}. For this, solving equation (\ref{2}), we have:
\begin{eqnarray}
y^{'}(z)=\frac{(\Sigma L)}{\sqrt{1-(\Sigma L)^{2}f(z)}}.\label{7}
\end{eqnarray}
In Fig.$\sim$\ref{p1}, we show the behavior of the equation (\ref{7}) for some values of the chemical potential with $\Sigma L=\cos(\theta)$ being $\theta$ the angle between the direction $y$ and the surface $Q$. In CFT, it is natural to say that the $\Delta y$ is the "width" of the boundary.
Taking the limit $\mu z\to0$ in the equation (\ref{7}) and considering the first term of this expansion, we obtain
\begin{eqnarray}
 y(z)=z\cot(\theta)\label{NB}.
\end{eqnarray}
$Q$ can be described as a plane intersecting the boundary $\mathcal{M}$ at an angle $\theta$. For the plane boundary, $A=\phi(z)dt$ depends on only the coordinate $z$.  For simplicity, we focus on the limit presented above, i.e., AdS$_{3}$ spacetime (\ref{3}) with the bulk boundary $Q$ located at (\ref{NB}). This limit is sufficient for the study of the leading order of current on the boundary. Moreover, Maxwell’s equations can be solved:
\begin{eqnarray}
\partial _{y}^{2}A_{t}+\partial _{z}^{2}A_{t}+\frac{1}{z}\partial_{z}A_{t}=0,\label{f1}
\end{eqnarray}
Similarly, the Neumann boundary on the gauge field $n^{\mu}\mathcal{F}_{\mu\nu}=0$, becomes
\begin{eqnarray}
(\partial_{y}A_{t}-\cot(\theta)\partial_{z}A_{t})|_{y=z\cot(\theta)}=0,\label{S2}
\end{eqnarray}
and analogously $\eta^{\mu}$$^{*}\mathcal{F}_{\mu\nu}=0$ \cite{Miao:2018qkc}. Following the prescription of \cite{Miao:2017aba,Chu:2018ntx}, we can consider the {\sl Ansatz} for the gauge field
\begin{eqnarray}
A_{t}(y,z)=A^{(0)}_{t}+A^{(1)}_{t}yf\left(\frac{z}{y}\right),\label{ansatz1}
\end{eqnarray}
 where $A^{(0)}_{t}$ and $A^{(1)}_{t}$ are constants. The Maxwell's equations for (\ref{f1}) become
\begin{eqnarray}
(1+s^{2})f^{''}(s)+\frac{1}{s}f^{'}(s)=0,
\end{eqnarray}
which has a general solution
\begin{eqnarray}
f(s)=1+b\sqrt{1+s^{2}}-b\ln\left(\frac{1}{s}+\sqrt{1+\frac{1}{s^{2}}}\right).
\end{eqnarray}
For the Neumann BC (\ref{S2}), we have
\begin{eqnarray}
b=\frac{2\frac{\cos^{2}(\theta/2)}{\cos(\theta)}-\frac{2}{\sin(2\theta)}}{\ln\left(2\frac{\cos^{2}(\theta/2)}{\sin(\theta)}\right)}.
\end{eqnarray}
Now let us go to consider the holographic theory. Thus, we consider the variation the on-shell action (\ref{1}) with respect to gauge fields, we have
\begin{eqnarray}
\delta I=\int_{\mathcal{M}}{d^{3}x\sqrt{-g}\eta_{\mu}F^{\mu\nu}\delta A_{\nu}}+\int_{Q}{d^{2}x\sqrt{-h}\eta_{\mu}F^{\mu\nu}\delta A_{\nu}}\label{ansatz2}.
\end{eqnarray}
On-shell, both $A_{\mu}$ on $\mathcal{M}$ and $Q$ are functions of background gauge fields $A_{i}$ of BCFT. However, based in \cite{Miao:2017aba,Chu:2018ntx}, from (\ref{ansatz1}) we have on $\mathcal{M}$
\begin{eqnarray}
A_{i}|_{\mathcal{M}}=\lim_{z\to 0}A_{i}=A^{(0)}_{i}+\sum^{\infty}_{n=1}{A^{(n)}_{i}y^{n}}=A_{i}\label{ansatz3},
\end{eqnarray}
where in a similar way, from $y(z)=z\cot(\theta)$ and (\ref{ansatz1}), we have on $Q$:

\begin{eqnarray}
A_{i}|_{Q}=A^{(0)}_{t}+A^{(1)}_{t}yf(\tan(\theta))\label{ansatz4}.
\end{eqnarray}
Now substituting (\ref{ansatz3}) and (\ref{ansatz4}) in (\ref{ansatz2}), we can derive the current via the definition

\begin{eqnarray}
J^{bdry}_{t}=\lim_{z\to 0}\frac{\delta I}{\delta A_{t}}=\partial^{2}_{z}A_{t}(y,z)=b_{1}A^{(1)}_{t}\label{ansatz5}.
\end{eqnarray}
Here we can see that near the boundary, the current in the equation (\ref{ansatz5}) with $b_{1}=-b\cot(\theta)$ playing the role of the central charge, which it is clear that the holographic BCFT satisfies the universal law of current \cite{Chu:2018ntx}. Besides, the current (\ref{ansatz5}) is independent of the $\Sigma$ near-boundary and consequently of boundary conditions.


\section{The entropy of the charged BTZ black hole}\label{v2}

In this section, we will compute the entropy for the BTZ black hole. For this, we need to execute the holographic renormalization scheme as presented by \cite{Takayanagi:2011zk,Santos:2021orr,Magan:2014dwa} to compute the Euclidean on-shell action, which is related to the free energy of the corresponding thermodynamic system, and then extract entropy. Thus, we start with the Euclidean action $I_{E}=I_{bulk}+2I_{boundary}$, i.e.,
\begin{eqnarray}
I_{bulk}=-\frac{1}{16\pi G_{N}}\int_{N}{d^{3}x\sqrt{g}\left((R-2\Lambda)-\frac{1}{4}F_{\mu\nu}F^{\mu\nu}\right)}-\frac{1}{8\pi G_{N}}\int_{\mathcal{M}}{d^{2}x\sqrt{\gamma}(K^{(\gamma)}-\Sigma^{(\gamma)})},
\end{eqnarray}
$\gamma$ and $\Sigma^{(\gamma)}$ are the induced metric and the surface tension on $\mathcal{M}$, respectively. $K^{(\gamma)}$ is the trace of the extrinsic curvature on the surface $\mathcal{M}$. On the other hand, for the boundary we have
\begin{eqnarray}
I_{bdry}=-\frac{1}{16\pi G_{N}}\int_{N}{d^{3}x\sqrt{g}\left((R-2\Lambda)-\frac{1}{4}F_{\mu\nu}F^{\mu\nu}\right)}-\frac{1}{8\pi G_{N}}\int_{Q}{d^{2}x\sqrt{h}(K-\Sigma)},
\end{eqnarray}
where $h$ and $\Sigma$, are respectively the induced metric and the surface tension on $Q$.

As we know from AdS/CFT correspondence, the IR divergences in the gravity side correspond to the UV divergences in the CFT boundary theory. Such relation is known as the IR-UV connection. Thus, for the AdS-BTZ black hole, we start computing the $I_{bulk}$, and we can note that by localizing the surface $\mathcal{M}$ in $z=0$ we can write $\Sigma^{(\gamma)}=1/L$, $K^{\gamma}=2/L$ with  
\begin{eqnarray} 
R=-\frac{6}{L^{2}}+\frac{\mu^{2}z^{2}}{2L^{2}},
\end{eqnarray}
and removing the IR divergence by introducing a cutoff $\epsilon$ the $I_{bulk}$ becomes   
\begin{eqnarray}
I_{bulk}&=&\frac{2z_{h}L\Delta y}{4G_{N}}\int^{z_{h}}_{\epsilon}{\frac{dz}{z^{3}}}-\frac{2z_{h}L\Delta y}{4G_{N}}\int^{z_{h}}_{z_{0}}{\frac{dz}{z^{3}}\left(\frac{\mu^{2}z^{2}}{8}\right)}\\
        &-&\frac{\mu^{2}kL^{4}\Delta y}{32z^{3}_{h}G_{N}}-\frac{1}{8\pi G_{N}}\int^{2\pi z_{h}}_{0}\int^{y_{0}}_{-y_{0}}{d\tau dy\frac{L\sqrt{f(\epsilon)}}{\epsilon^{2}}},\nonumber\\
        &=&-\frac{Lz_{h}\Delta y}{4G_{N}}\left(\frac{1}{z^{2}_{h}}-\frac{1}{\epsilon^{2}}\right)-\frac{Lz_{h}\mu^{2}\Delta y}{16G_{N}}\ln\left(\frac{z_{h}}{\epsilon}\right)-\frac{\mu^{2}L^{4}\Delta y}{32z^{3}_{h}G_{N}}-\frac{Lz_{h}\Delta y}{4G}\frac{\sqrt{f(\epsilon)}}{\epsilon^{2}}.\nonumber
\end{eqnarray}
Note that for a power expansion in $\epsilon$ we can write
\begin{eqnarray}
I_{bulk}=-\frac{L\Delta y}{4z_{h}G_{N}}-\frac{\mu^{2}L^{4}\Delta y}{32z^{3}_{h}G_{N}}-\frac{Lz_{h}\mu^{2}\Delta y}{16G_{N}}\ln\left(\frac{z_{h}}{\epsilon}\right)+\mathcal{O}(\epsilon),
\end{eqnarray}
Now, for the $I_{bdry}$ with $K=2\Sigma-\Sigma\mu^{2}z^{2}$, we have
\begin{eqnarray}
&&I_{bdry}=-\frac{L\Delta y}{4z_{h}G_{N}}-\frac{\mu^{2}L^{4}\Delta y}{32z^{3}_{h}G_{N}}-\frac{Lz_{h}\mu^{2}\Delta y}{16G_{N}}\ln\left(\frac{z_{h}}{\epsilon}\right)-\frac{L}{4G_{N}}arc\sinh(\cot(\theta))-\nonumber\\
&&\frac{\mu^{2}z^{4}_{h}}{8(\Sigma L)^{3}G_{N}}\left[\Sigma L+(1-(\Sigma L)^{2})\ln\left(\frac{\sqrt{1-(\Sigma L)^{2}}}{1+\Sigma L}\right)\right]+\mathcal{O}(\epsilon),
\end{eqnarray}
Then using the relation $I_{E}=I_{bulk}+2I_{bdry}$, we have
\begin{eqnarray}
I_{E}=-\frac{L\Delta y}{4z_{h}G_{N}}-\frac{L}{2G_{N}}arc\sinh(\cot(\theta))-\frac{\Phi(\mu,z_{h})}{4G_{N}},
\end{eqnarray}
where the function $\Phi(\mu,z_{h})$ is of the form
\begin{eqnarray}
&\Phi(\mu,z_{h})=\frac{\mu^{2}L^{4}\Delta y}{2z^{3}_{h}}-\frac{\mu^{2}\Delta yz^{4}_{h}}{(\Sigma L)^{3}}\left[\Sigma L+(1-(\Sigma L)^{2})\ln\left(\frac{\sqrt{1-(\Sigma L)^{2}}}{1+\Sigma L}\right)\right],
\end{eqnarray}
and we have the entropy as
\begin{eqnarray}
S=-\frac{\partial F}{\partial T}=-\frac{\partial(T_{H}I_{E})}{\partial T_{H}},
\end{eqnarray}
that is,
\begin{eqnarray}
S=\frac{L\Delta y}{8z_{h}G_{N}}+\frac{L}{G_{N}}arc\sinh(\cot(\theta))+\frac{\tilde{\Phi}(\mu,z_{h})}{4G_{N}}\label{SB},
\end{eqnarray} 
with
\begin{eqnarray}
&&\tilde{\Phi}(\mu,z_{h})=\frac{4-\mu^{2}z^{2}_{h}}{4+\mu^{2}z^{2}_{h}}\left(\frac{L\Delta y}{2z_{h}}+\frac{3\mu^{2}L^{4}\Delta y}{2z^{3}_{h}}-\frac{\mu^{2}z_{h}L\Delta y}{2}\right)+\nonumber\\
&&\frac{4-\mu^{2}z^{2}_{h}}{4+\mu^{2}z^{2}_{h}}\left(-\frac{4\Delta y\mu^{2}z^{4}_{h}}{(\Sigma L)^{3}}\left[\Sigma L+(1-(\Sigma L)^{2})\ln\left(\frac{\sqrt{1-(\Sigma L)^{2}}}{1+\Sigma L}\right)\right]\right),
\end{eqnarray}
Equation (\ref{SB}) is the total entropy of the AdS-BTZ black hole with electric field contribution terms for the bulk and the boundary $Q$. This entropy does not respect the Bekenstein-Hawking (BH) scaling, i.e., it is not equal to the area of the "`shadow" of the bounding surface $Q$ on the horizon (marked orange in figure \ref{p1}) times the correct BH factor. Furthermore, still analyzing Eq. (59), due to the limit $T\to0$ ($z_{h}\to\infty$) in Eq. (\ref{SB}); then, one gets the $z_{h}=2/\mu$. Our boundary entropy becomes
\begin{eqnarray}
S_{Q}=\frac{L}{G_{N}}arc\sinh(\cot(\theta))\label{SB1}.
\end{eqnarray} 
where $S_{Q}$ in equation (\ref{SB1}) is the boundary entropy. This entropy is carried by the boundary and increases as the tension does. On the other hand, at the high temperature phase with $T\to\infty$ ($z_{h}\to0$), we have, $T=T_{BCFT}=1/2\pi z_{h}$ and the equation (\ref{SB}) can be written as
\begin{eqnarray}
S=S_{bulk}+S_{Q}=\frac{L\Delta y}{8z_{h}G_{N}}\left(1+\frac{3\mu^{2}L^{3}}{2z^{2}_{h}}\right)+\frac{L}{G_{N}}arc\sinh(\cot(\theta))\label{SB2},
\end{eqnarray} 
A similar result to the bulk entropy ($S_{bulk}$) in equation (\ref{SB2}) was presented by \cite{Hartnoll:2009sz}. However, in our case, the leading nontrivial temperature dependence of the entropy ($S_{bulk})$ , as is the leading high temperature dependence of the heat capacity ($C_{V}$)
\begin{eqnarray}
C_{V}=T\left(\frac{\partial S}{\partial T}\right)_{V}\label{SB3},
\end{eqnarray}
the above equation is a positive quantity,  implying that the BTZ black hole within an electric field at a high temperature is stable. In the next section, we will show the consistency of thermodynamic relations of the fluid, described by $T_{ab}$ on $Q$.  

\section{Boundary fluid from AdS$_{3}$/BCFT$_{2}$}\label{v3}

\begin{figure}[!ht]
\begin{center}
\includegraphics[scale=0.6]{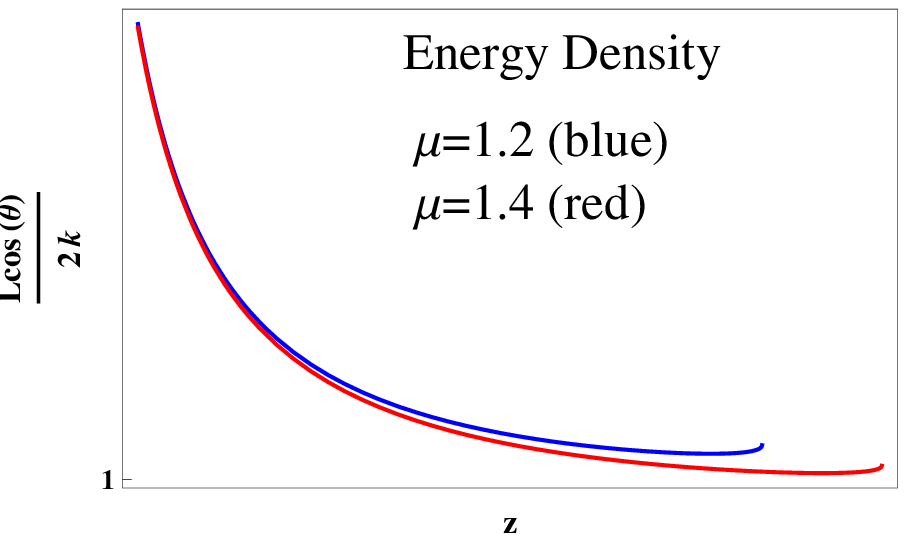}
\includegraphics[scale=0.6]{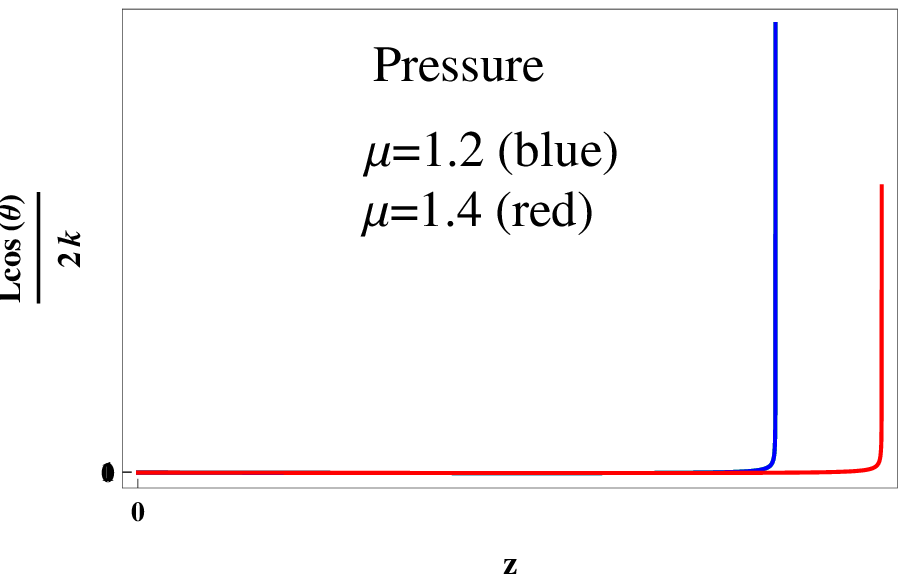}
\includegraphics[scale=0.6]{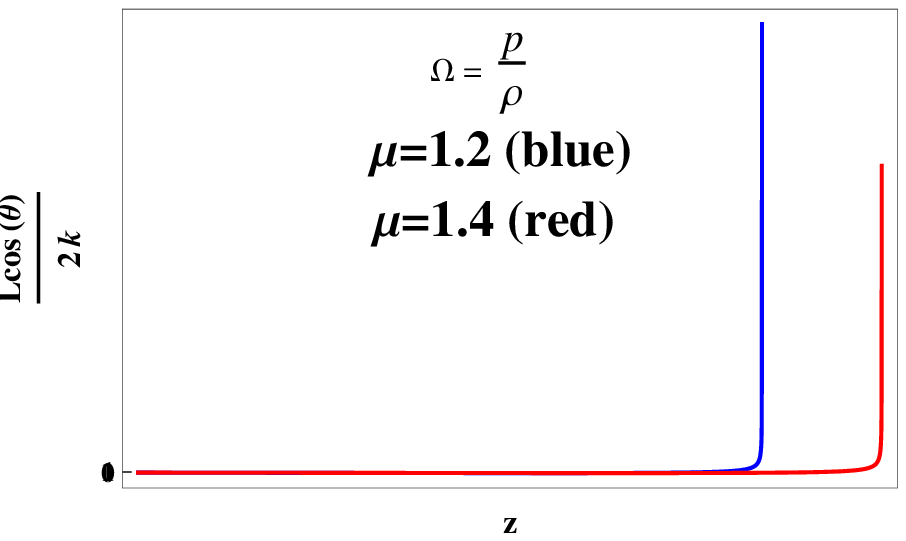}
\caption{The figure shows the behavior of the energy density (\ref{j1}) and pressure (\ref{j2}) within an electric field where the equation of state (\ref{rho3}) provides an effective theory on the surface $Q$.}\label{pr}
\label{planohwkhz}
\end{center}
\end{figure}

In this section we apply the discussion of \cite{Magan:2014dwa} to fluid/gravity correspondence in the AdS$_{3}$/BCFT$_{2}$ correspondence. In our case, we have that for $d=3$, the boundary stress tensor is given by
\begin{eqnarray}
T_{ab}=-\frac{L}{kz}(K_{ab}-Kh_{ab}+\Sigma h_{ab}-kT^{R}_{ab}-kT^{ct}_{ab}),\label{Tensor}
\end{eqnarray}
where $k=8\pi G$. $T^{R}_{ab}$ and $T^{ct}_{ab}$ are, respectively, possible contributions of the intrinsic curvature and counter-terms. But with the energy-momentum tensor fixed on the surface $Q$ with $T^{R}_{ab}=T^{ct}_{ab}=0$, the equation (\ref{Tensor}) can be written as
\begin{eqnarray}
T_{ab}=-\frac{L}{kz}(K_{ab}-Kh_{ab}+\Sigma h_{ab}).\label{Tensor1}
\end{eqnarray}
Using the equation (\ref{Tensor1}), we will show that there is a single profile among a continuum of the profiles of $Q$ (or stress-energy tensors $T_{ab}$). In this way, gravity yields a fluid-like stress-energy tensor, in local thermodynamic equilibrium for the BTZ black hole within the electric field. In the AdS$_{3}$/BCFT$_{2}$ correspondence, we parametrize the surface $Q$ by the function $y(z)$ with the arbitrary tension $\Sigma$. The energy density and pressure are given by:
\begin{eqnarray}
\rho&=&\frac{L}{kz}\left(\Sigma L+\frac{zy^{'}f^{'}+2zfy^{''}-2f^{2}y^{'3}-2y^{'}f}{2(1+y^{'2}f)^{3/2}}\right),\label{j01}\\
p_{yy}&=&-\frac{L}{kz}\left(\Sigma L-\frac{2(2f-zf^{'})y^{'}+f(4f-zf^{'})y^{'3}-2zfy^{''}}{2(1+y^{'2}f)^{3/2}}\right),\label{jo2}\\
p_{zz}&=&-\frac{L}{kz}\left(\Sigma L-\frac{4y^{'}f-zy^{'}f^{'}}{2(1+y^{'2}f)^{1/2}}\right).\label{jo3}
\end{eqnarray}
The above equations show that there is a symmetry break through the pressure components, i.e., $p_{yy}\neq p_{zz}$ and the space is anisotropic. If we want a system that describes a Pascal fluid-like with $p_{yy}=p_{zz}$, this condition implies that $fy^{'2}=c$, where $c$ is a constant. Thus, we have
\begin{eqnarray}
y-y_{0}=\int^{z}_{0}{\frac{\cot(\theta)d\omega}{\sqrt{f(\omega)}}},
\end{eqnarray}
where $c=\cot^{2}(\theta)$ and using the condition $fy^{'2}=c$, we can write energy density and pressure as:
\begin{eqnarray}
\rho&=&\frac{2L\cos(\theta)}{kz}\left(1-\frac{\sqrt{f}}{2}\right),\label{j1}\\
p&=&\frac{L\cos(\theta)}{2kz\sqrt{f}}(4f-zf^{'}-4\sqrt{f}),\label{j2}
\end{eqnarray}
The equation of state for the energy density and pressure can be written as
\begin{eqnarray}
\Omega=\frac{p}{\rho}=\frac{4f(z)-zf^{'}(z)-4\sqrt{f(z)}}{\sqrt{f(z)}(2-\sqrt{f(z)})}\label{rho3}.
\end{eqnarray}
In our fluid beyond the gravitational force, due to the non-flat space, we have an external electric field that decreases the energy density and increases the pressure, see figure \ref{pr}. This effect is captured by a confining aspect of the Maxwell theory encoded in the logarithmic behavior of the charge interaction potential. However, taking the regime $z\rightarrow 0$, this regime is very important to see if the fluid has a conformal behavior. Thus, we can to extract the following $\rho=L\cos(\theta)/\kappa z_{h}$, $p=-\mu^{2}z_{h}L\cos(\theta)/4\kappa$ and $\Omega=-2\mu^{2}z^{2}_{h}$. Now, with these relations, we can conclude that $p+\rho=sT$ where:
\begin{eqnarray}
s_{Q}=\frac{2\pi L}{\kappa}\cos(\theta),
\end{eqnarray}
The entropy density is coordinate independent. In low temperature $T\to0$ ($z_{h}\to\infty$), we have $p=-\rho$, i.e, the fluid is non-conformal. Even with the $Q$ profile open in the direction of the horizon, which is a consequence of a "null (weak)" energy condition in $Q$. Where this condition is required for the temperature or energy density in $Q$ to be non-negative. We can note that the electric field breaks the conformal symmetry in this case. However, in the regime of high temperature $T\to\infty$ ($z_{h}\to0$), we have $T=T_{BCFT}=1/2\pi z_{h}$ and the electric field non-broken the conformal symmetry.

\newpage
\section{Conclusions}\label{v4}

In this section we present the conclusions on the AdS/BCFT correspondence and BTZ black hole within the electric field. Considering the coupling with the Maxwell field and the Einstein tensor, we established our setup. In the three-dimensional bulk, we found a consistent solution for the BTZ black hole within the electric charge. From these solutions, we constructed the $Q$ profile on the two-dimensional boundary, which characterizes the AdS$_{3}$/BCFT$_{2}$ correspondence. In particular, we found a numerical solution  to the $Q$ profile, which is open in the direction of the horizon where we have a "null (weak)" energy condition in $Q$ with a positive temperature and energy density in $Q$. Such a solution described qualitatively well the influence of the electric field. On the other hand, when we consider the limit near the boundary, the solution is sufficient to compute the first order of the electric current in the boundary, it satisfies the universal law of current \cite{Chu:2018ntx}. Performing the holographic renormalization procedure to get the Euclidean on-shell action for the AdS-BTZ black hole within the electric field, we identify that the Euclidean on-shell action with the free energy. With this, we compute the total entropy and discuss the low and high temperature regimes. Such regimes are compatible with those expected from usual black hole thermodynamic properties.

In the fluid/gravity duality, we noticed a special solution that yields the same "conformal" fluid of the standard fluid/gravity calculation in the AdS-BTZ black hole geometry, albeit with thermodynamic quantities dependent on position (inhomogeneous fluid). For local quantities, we show that it describes this boundary fluid. Specifically, we have found that the entropy density does not depend on the coordinates.

\section*{Acknowledgment}

We would like to thank CNPq and CAPES for partial financial support.

\end{document}